\documentclass[aps,twocolumn,prl,showpacs,eqsecnum]{revtex4}

\usepackage[dvips]{graphicx}

\begin{document}

\title{Local pressure-induced metallization of a semiconducting carbon nanotube in a crossed junction}

\author{L.Vitali, M.Burghard, P.Wahl, M.A.Schneider, K.Kern}
\address{Max-Planck-Institut f\"ur Festk\"orperforschung,
Heisenbergstr.1, D-70569 Stuttgart, Germany}

\date{\today}

\begin{abstract}
The electronic and vibrational density of states of a
semiconducting carbon nanotube in a crossed junction was
investigated by elastic and inelastic scanning tunneling
spectroscopy. The strong radial compression of the nanotube at the
junction induces local metallization spatially confined to a few
nm. The local electronic modifications are correlated with the
observed changes in the radial breathing and G-band phonon modes,
which react very sensitively to local mechanical deformation. In
addition, the experiments reveal the crucial contribution of the
image charges to the contact potential at nanotube-metal
interfaces.
\end{abstract}

\maketitle

Single-wall carbon nanotubes (SWCNTs) are highly attractive
components in nano-scale electronics such as high-performance
field-effect transistors \cite{heinze03} or electrical
interconnects \cite{kong01}. The successful implementation of
advanced device functions into nanotubes will critically depend on
the availability of methods that allow to tune the tubes'
electronic properties in a well-defined and spatially controllable
manner. One possibility to reach this goal is to use
conformational changes, which are well-documented to alter the
electronic properties of the nanotubes. Such changes can be
induced locally by e.g. mechanical interactions. When a nanotube
encounters topographic obstacles like steps in the substrates,
electrode lines or other nanotubes, deformations like bends or
kinks are introduced into the tube
 which can result in the local
formation of tunnel barriers \cite{hertel98,bezryadin98} or other
electronic modifications \cite{mazzoni00, nardelli99}.
Alternatively, the tip of an atomic force microscope can be used
to alter the electronic structure of a nanotube via mechanical
manipulation with spatial control in the nanometer range
\cite{postma01}. In order to take full advantage of nanomechanical
interactions to control the electronic properties of nanotubes and
thereby tailoring nanodevices, a detailed microscopic
understanding of mechanical nanotube contacts is mandatory.
\par\
Particularly favorable configurations to study these contacts are
crossed junctions between nanotubes. The interaction between the
tubes is well defined and spatially localized. Theoretical
calculations predict a delicate interplay between the structural
deformation and the electronic properties of the nanotube junction
\cite{mazzoni00, nardelli99} with rich physics ranging from band
gap modifications through the formation of localized states to
metal-semiconductor transitions. Experimentally crossed nanotube
junctions can be addressed by scanning tunneling microscopy, but
very few studies have been reported so far \cite{janssen02}.  Here
we report on the simultaneous experimental determination of the
local changes in the electronic and vibrational properties in an
individual semiconducting SWCNT induced by its bending over a thin
SWCNT bundle. The electronic and vibrational response to local
mechanical distortions is mapped with atomic resolution via
elastic and inelastic tunneling spectroscopy \cite{ vitali04b}.
Both, electronic and phonon density of states are found to react
very sensitively to local tube deformations. Most remarkably, we
demonstrate a local metallization of the semiconducting nanotube
at the crossing point.

In our experiments, as-produced HiPco (tubes@Rice) SWCNTs have
been ultrasonically dispersed in 1,2-dichloroethane and deposited
on Au/mica substrates. The samples have been transferred into an
UHV system equipped with a home-built STM, and tunneling
experiments performed at 6~K using an iridium tip. Recording the
first and the second derivative of the tunneling current allows
for spatial mapping of the electronic and vibrational density of
states with atomic resolution \cite{vitali04b}

\begin{figure}[ht]
 \centering
   \includegraphics[draft=false,keepaspectratio=true,clip,%
                   width=0.9\linewidth]%
                   {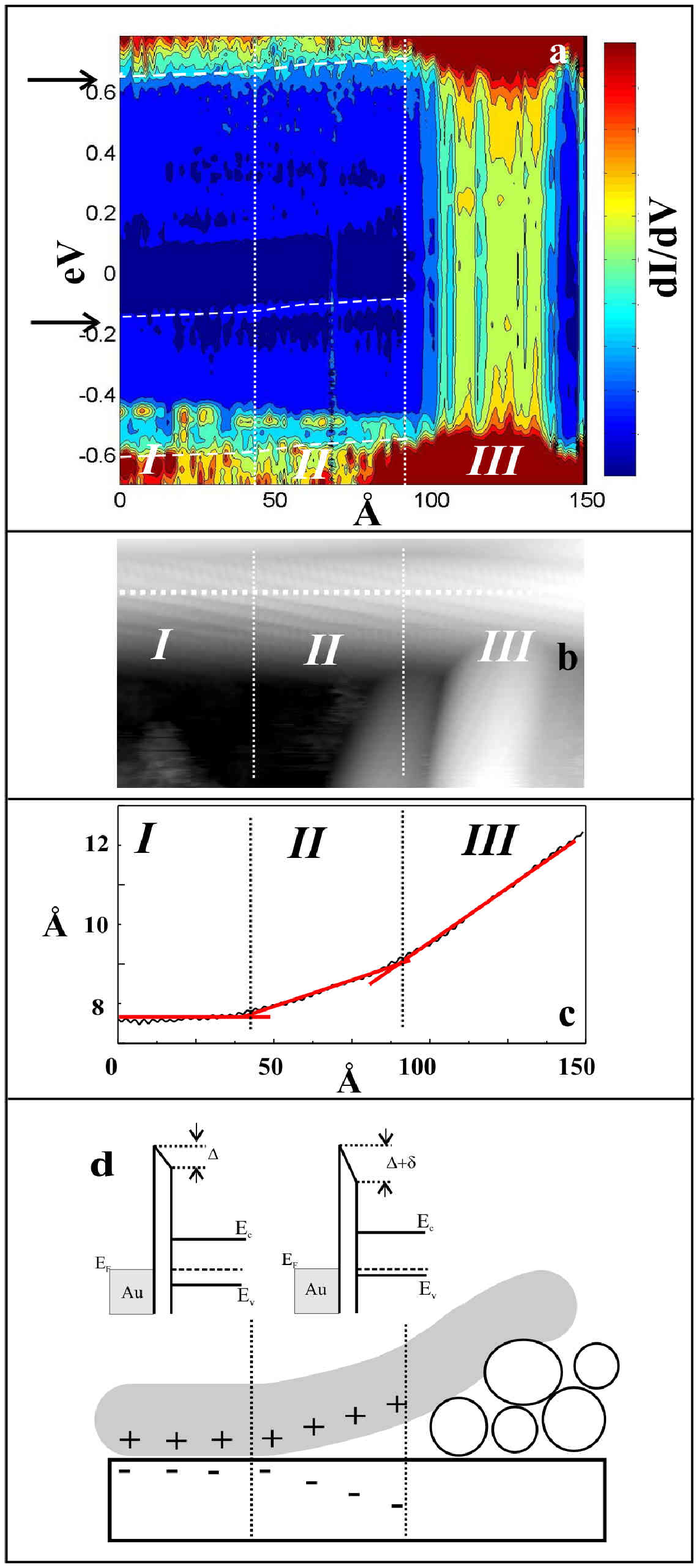}

 \caption{\label{fig1}  a) Local electronic density of states of an (9,2) carbon nanotube crossing a bundle, as revealed by
 elastic scanning tunneling spectroscopy.The dI/dV signal is measured under open feedback conditions
 along the tube axis. b) Topographic STM image of the
 corresponding nanotube; note that the tube extends for further
 150 Å on the right side of the crossing point.
  c) Line profile along the tube axis showing the three different regions.
  d) Schematic description of the tube and the corresponding contact potential.}
 \label{samplefig}
\end{figure}

A topographic image of the investigated SWCNT crossing a bundle,
with an apparent height of 14\AA, is depicted in figure
\ref{fig1}b. To identify this tube, the procedure described by
Wild{\"o}er \cite{wildoer98} was followed. Using the energy
distance between the first van Hove singularities (0.88eV,
\ref{fig1}a), combined with the chiral angle obtained from the
topographic image (fig.\ref{fig1}b), allows the assignment to the
semiconducting (9,2) tube. The line profile, shown in figure
\ref{fig1}c, indicates that the tube does not follow closely the
contour of the bundle below. Three regions can be distinguished
along the tube: in region (I) the tube lies directly on the gold
surface, while in region (II) it is lifted from the substrate and
suspended between the Au surface and the SWCNT bundle. Finally, in
region (III) the tube starts to bend over the bundle. The
corresponding topography is schematically illustrated in figure
\ref{fig1}d.
\par\
Within region (III), which contains the intersection point between
the (9,2) tube and the bundle, a striking change in the electron
density of states is recognized (figure \ref{fig1}a). The
conductivity at the Fermi level is considerably enhanced, and the
original energy gap no longer visible, indicating that the
nanotube assumes metallic character in this region. Shortly after
the crossing point, the initial semiconducting character of the
(9,2) tube is restored, which demonstrates the very local
character of the metallization. While previous work reported
strong perturbations of the electron density of states and the
formation of tunneling barriers as a consequence of tube crossings
or interaction with patterned substrates
\cite{fuhrer00,jassen02,bezryadin98}, the local metallization of a
semiconducting tube has not yet been observed.

Before discussing in detail this local metallization we first
analyze the electronic and vibrational structure of the adsorbed
and freely suspended tube segments. Within the suspended region
(II), it can be seen that the van Hove singularities in the
valence and conduction bands experience a shift of approximately
60meV towards higher energy with respect to their position in
region (I).  This shift occurs mostly within the first 3nm of
region (II), which is followed by a much weaker dependence with
increasing distance from the substrate. The nanotube-metal
substrate interface has been subject of several theoretical
investigations. According to these, the difference in
work-function between the two materials leads to charge transfer
across the metal-tube interface. The resulting charge accumulation
layer  in the nanotube and its image-charge in the metallic
substrate give rise to a contact potential
$\Delta=\Delta_{Holes}-\Delta_{Images}$ at the interface which
affects the alignment of the nanotube bands, causing their
displacement towards higher energies \cite{xue99,czerw02}. The
charge density at the interface decays only logarithmically with
the distance \cite{leonard99,odintsov00}, due to the ineffective
Coulomb screening in a quasi one-dimensional nanotube. Hence, in
first approximation, the charge density in the nanometer-long
suspended segment can be assumed  as constant. The increasing
distance to the substrate weakens, however, the screening effect
of the image-charges (figure \ref{fig1}c), leading to an apparent
excess charge on the nanotube and thus to an increased contact
potential at the interface $\Delta'=\Delta+\delta$, which shifts
the bands further towards higher energies (figure \ref{fig1}c).

The modified capacitive coupling of the substrate-nanotube-tip
system, in the suspended region, can be addressed in analogy to
double barrier tunneling junctions \cite{hanna92}. Using the
relation ${\Delta}$V=${\Delta}$Q/C$_{TS}$ where ${\Delta}$V is the
voltage shift, ${\Delta}$Q the fractional charge at the interface
and C$_{TS}$ the tube-substrate capacitance, the unscreened charge
can be estimated from the measured shift of the valence band
singularity. For the capacitance C, we apply the model of a finite
wire of length L at a distance z from a conducting plate
$C=2\pi\epsilon_0\epsilon_rL/\mathrm{ln}(4z/d)$ \cite{springer80}.
Considering the measured energy shift of 60mV, an averaged
tube-substrate distance of 1nm,  a length of 3nm for the tube
section within which most of the shift occurs, and assuming
$\epsilon_r$ =2 , the reduced screening is associated with an
additional charge of ~0.08 holes on the suspended tube segment.
Noting that theoretical predictions estimate that, upon bringing
the nanotube into contact with a gold surfaces, 0.3 holes are
transferred in the same tube length \cite{rubio99}, the reduced
screening of the image-charge is equivalent to an additional
26$\%$ of the total charge transferred at the interface. This
result suggests that the contact potential at the tube-metal
interface is determined not only by the difference in the work
functions, but also significantly influenced by the image-charge.
This property places the nanotube interface at an intermediate
position between the bulk inorganic semiconductor-metal interface,
where the image-charge influences only marginally the contact
barrier, and the organic semiconductor-metal interface, where the
contribution of the image-charge is comparable to the built-in
potential \cite{knupfer04}.

\begin{figure}[ht]
 \centering
   \includegraphics[draft=false,keepaspectratio=true,clip,%
                   width=1\linewidth]%
                   {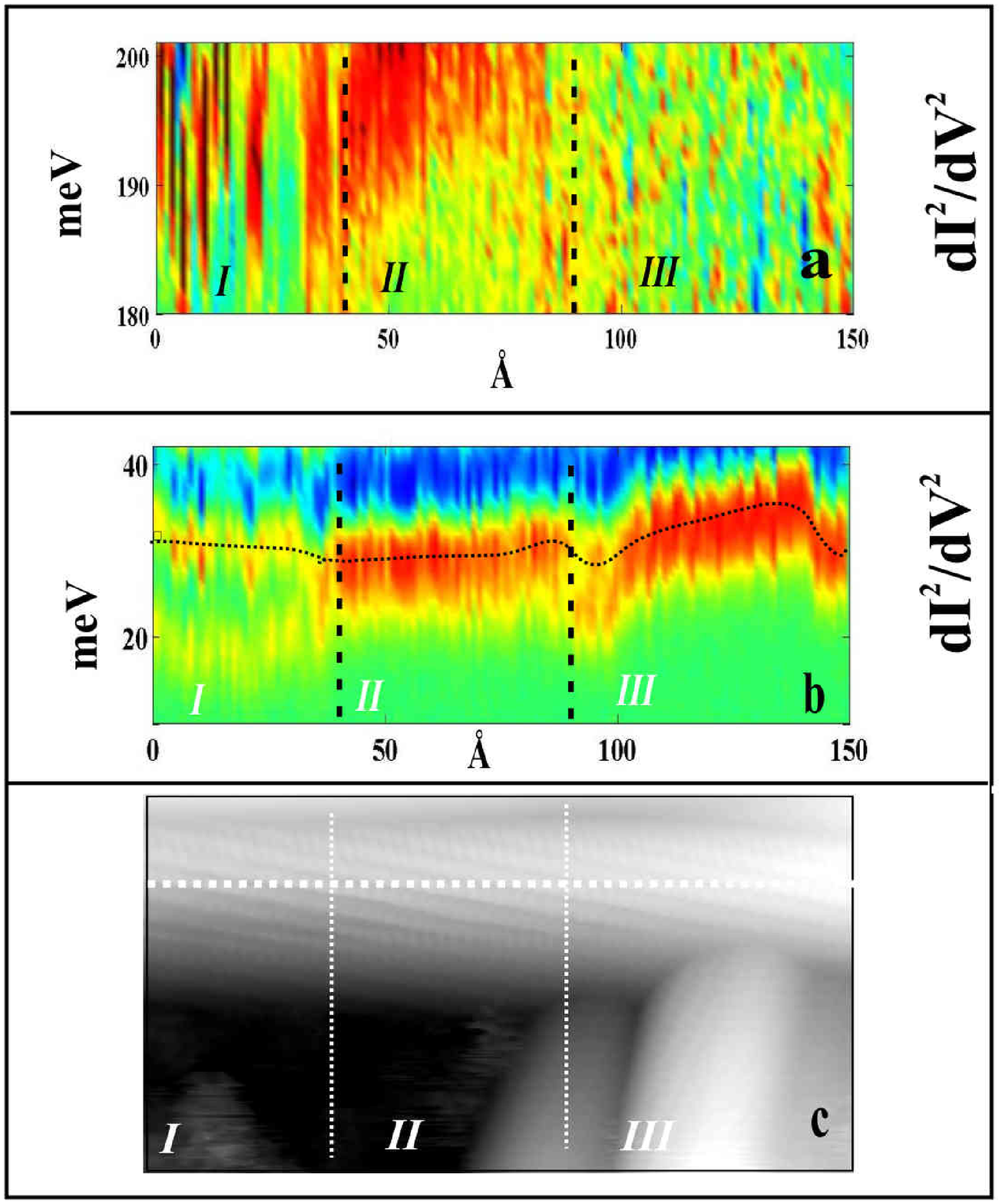}
 \caption{\label{fig2} Local vibrational density of states of the (9,2) carbon nanotube crossing a bundle as revealed by
inelastic scanning tunneling spectroscopy. the $dI^{2}/dV^{2}$
signal is measured in open feedback
 conditions along the tube axis. The region of the tangential mode
 a) and the radial breathing mode b) are displayed; the dotted
 line (in b)is a guide to the eye.
 c) Corresponding STM topography.}
 \label{samplefig}
\end{figure}

The behavior of the most prominent phonon modes, namely the radial
breathing mode (RBM) and the G-band apparent from the
corresponding IETS maps (figure \ref{fig2}), which display the
total phonon density of states as a function of position along the
tube axis. In view of the low sensitivity of the G-band and of the
RBM to charge doping \cite{corio03,gupta04},  it is not expected
that the small additional charge in the tube segment of region
(II) ($\sim 1\times10^{-4}$ holes per carbon atom) shifts these
modes significantly.   As can be seen in figure \ref{fig2}a, the
G-band, which consists of 6 unresolved vibrational modes
\cite{dresselhaus00} corresponding to lattice displacements in the
circumferential direction (tangential modes) and along the tube
axis (longitudinal modes), occurs in region (I) between 190 and
200 meV. Within region (II), the energy position of the
longitudinal modes stays constant while a gradual decrease of the
intensity of the tangential displacements is observed. A similar
observation has been made in case of chemically doped carbon
nanotubes \cite{dresselhaus00}. The RBM, which comprises the
in-phase radial displacement of the carbon atoms and whose energy
scales inversely with the tube diameter occurs on the metal
substrate at an energy of ~32meV (see figure \ref{fig2}b), in good
accordance to the value expected for the (9,2) tube with a
diameter of 8.1\AA\ \cite{vitali04b}. Within region (II), the RBM
experiences a slight downshift by about 1.5meV as compared to
region (I). This downshift can be attributed to the reduced
interaction forces with the substrate as a result of lifting the
tube. The importance of the surrounding is well-documented in
nanotube bundles, where van-der-Waals interactions cause
stiffening of the RBM \cite{henrard99}. For the (9,2) tube, the
RBM is expected to harden by 6$\%$ due to these interactions with
respect to the free-standing tube. Although the van-der-Waals
interaction of the tube with the surface is not isotropic as in a
nanotube bundle, the experimentally observed decrease in the RBM
energy of  1.5meV upon entering region (II), is largely consistent
with the estimated value of ~1.9meV.
\par\

The metallization observed in region (III) is ascribed to a local
compression of the tube. The van der Waals forces attracting the
upper tube to the substrate result in a large local pressure at
the crossing region. This causes an elliptical shape of the tube
as can be deduced from the line profile (figure \ref{fig1}c)
showing a height that is about 35$\%$ smaller than the sum of the
apparent heights of the tube and the crossed bundle.  The
compressibility of the tube, which is described by the bulk
modulus B, provides for an estimate of the pressure P imposed at
the crossing junction through the infinitesimal relation
dP=-dV/VB, where V is the volume. Assuming that the attractive
van-der-Waals forces between the upper nanotube and the surface
squeeze the tube cross-section into an elliptical shape at the
crossing point, and using a bulk modulus of 35GPa \cite{sharma01},
the pressure necessary to obtain a volume reduction of 35$\%$ is
estimated to 15GPa. This pressure is in reasonable agreement with
that predicted by first principle calculations \cite{hertel98} for
nanotubes at crossing-junctions. It is remarkable that for the
(9,2) tube, an elliptical cross section with an axis ratio of 0.65
implies a short axis length that is very close to the
theoretically predicted 4.6\AA\ necessary to induce a
semiconductor-metal transition \cite{mazzoni00}. It is therefore
likely that the observed tube metallization is induced by the
local structural change. The strong deformation of the nanotube at
the crossing junction is reflected in the behavior of the phonon
density of states in region (III). In particular, the G-band is
completely suppressed within region (III) (figure \ref{fig2}a).
This observation agrees well with results from high pressure Raman
studies, which demonstrated a strong attenuation of the G-band
above 10GPa and its complete vanishing at 20GPa as a consequence
of the strong radial compression of the SWCNTs \cite{loa03}. The
energy of the RBM rises to the value of 37meV at the crossing
point (figure \ref{fig2}b) and quickly restores afterwards. The
observed energy variation of the RBM between the suspended and the
compressed region ($\sim$7 meV) is lower than the value expected
from Raman investigations reporting an increase by 1.1meV/GPa.
However, it is noticed that the present study addresses effective
pressures well above 2GPa, which represents the maximum that could
so far been investigated by Raman spectroscopy due to the loss of
the resonance enhancement. Hence, it cannot be excluded that the
energy vs. pressure dependence of the RBM would assume a
significantly weaker slope in the high pressure regime, or that
the non-hydrostatic pressure affects the RBM in a different
manner. In figure \ref{fig2}b, one furthermore observes a "wiggle"
(local minimum) in the energy dependence of the RBM at the left
border of region (III), where the first and the second tube of the
bundle are touching. This behaviour further illustrates the
extreme sensitivity of nanotube phonon modes to local mechanical
interactions.
\par\

In conclusion, the combination of elastic and inelastic scanning
tunneling spectroscopy techniques has revealed a correlation
between changes of the local electron and phonon density of states
of a semiconducting SWCNT, induced by the bending of the tube over
a thin nanotube bundle. In the suspended tube region, the observed
shift of the valence band towards the Fermi energy has been
attributed to an effective hole doping resulting from reduced
screening of the charge accumulation layer by the metal substrate.
Around the crossing point, the tube achieves a metallic character.
This striking observation has been attributed to the strong local
radial compression, which is corroborated by the observed increase
in the RBM energy as well as the complete disappearance G-band in
this tube region.

\bibliography{mio}

\begin{thebibliography}{26}
\expandafter\ifx\csname natexlab\endcsname\relax\def\natexlab#1{#1}\fi
\expandafter\ifx\csname bibnamefont\endcsname\relax
  \def\bibnamefont#1{#1}\fi
\expandafter\ifx\csname bibfnamefont\endcsname\relax
  \def\bibfnamefont#1{#1}\fi
\expandafter\ifx\csname citenamefont\endcsname\relax
  \def\citenamefont#1{#1}\fi
\expandafter\ifx\csname url\endcsname\relax
  \def\url#1{\texttt{#1}}\fi
\expandafter\ifx\csname urlprefix\endcsname\relax\def\urlprefix{URL }\fi
\providecommand{\bibinfo}[2]{#2}
\providecommand{\eprint}[2][]{\url{#2}}

\bibitem[{\citenamefont{S.Heinze et~al.}(2003)\citenamefont{S.Heinze,
  J.Tersoff, and P.Avouris}}]{heinze03}
\bibinfo{author}{\bibnamefont{S.Heinze}},
  \bibinfo{author}{\bibnamefont{J.Tersoff}}, \bibnamefont{and}
  \bibinfo{author}{\bibnamefont{P.Avouris}}, \bibinfo{journal}{Applied Physics
  Letters} \textbf{\bibinfo{volume}{83}}, \bibinfo{pages}{5038}
  (\bibinfo{year}{2003}).

\bibitem[{\citenamefont{J.Kong et~al.}(2001)\citenamefont{J.Kong, E.Yenilmez,
  T.W.Tombler, W.Kim, H.J.Dai, R.B.Laughlin, L.Liu, C.S.Jayanthi, and
  S.Y.Wu}}]{kong01}
\bibinfo{author}{\bibnamefont{J.Kong}},
  \bibinfo{author}{\bibnamefont{E.Yenilmez}},
  \bibinfo{author}{\bibnamefont{T.W.Tombler}},
  \bibinfo{author}{\bibnamefont{W.Kim}},
  \bibinfo{author}{\bibnamefont{H.J.Dai}},
  \bibinfo{author}{\bibnamefont{R.B.Laughlin}},
  \bibinfo{author}{\bibnamefont{L.Liu}},
  \bibinfo{author}{\bibnamefont{C.S.Jayanthi}}, \bibnamefont{and}
  \bibinfo{author}{\bibnamefont{S.Y.Wu}}, \bibinfo{journal}{Physical Review
  Letters} \textbf{\bibinfo{volume}{87}}, \bibinfo{pages}{106801}
  (\bibinfo{year}{2001}).

\bibitem[{\citenamefont{T.Hertel et~al.}(1998)\citenamefont{T.Hertel,
  R.E.Walkup, and P.Avouris}}]{hertel98}
\bibinfo{author}{\bibnamefont{T.Hertel}},
  \bibinfo{author}{\bibnamefont{R.E.Walkup}}, \bibnamefont{and}
  \bibinfo{author}{\bibnamefont{P.Avouris}}, \bibinfo{journal}{Physical Review
  B} \textbf{\bibinfo{volume}{58}}, \bibinfo{pages}{13870}
  (\bibinfo{year}{1998}).

\bibitem[{\citenamefont{A.Bezryadin et~al.}(1998)\citenamefont{A.Bezryadin,
  A.R.M.Verschueren, S.J.Tans, and C.Dekker}}]{bezryadin98}
\bibinfo{author}{\bibnamefont{A.Bezryadin}},
  \bibinfo{author}{\bibnamefont{A.R.M.Verschueren}},
  \bibinfo{author}{\bibnamefont{S.J.Tans}}, \bibnamefont{and}
  \bibinfo{author}{\bibnamefont{C.Dekker}}, \bibinfo{journal}{Physical Review
  Letters} \textbf{\bibinfo{volume}{80}}, \bibinfo{pages}{4036}
  (\bibinfo{year}{1998}).

\bibitem[{\citenamefont{M.S.C.Mazzoni and H.Chacham}(2000)}]{mazzoni00}
\bibinfo{author}{\bibnamefont{M.S.C.Mazzoni}} \bibnamefont{and}
  \bibinfo{author}{\bibnamefont{H.Chacham}}, \bibinfo{journal}{Applied Physics
  Letters} \textbf{\bibinfo{volume}{76}}, \bibinfo{pages}{1561}
  (\bibinfo{year}{2000}).

\bibitem[{\citenamefont{Nardelli and J.Bernholc}(1999)}]{nardelli99}
\bibinfo{author}{\bibfnamefont{M.}~\bibnamefont{Nardelli}} \bibnamefont{and}
  \bibinfo{author}{\bibnamefont{J.Bernholc}}, \bibinfo{journal}{Physical Review
  B} \textbf{\bibinfo{volume}{60}} (\bibinfo{year}{1999}).

\bibitem[{\citenamefont{H.W.C.Postma et~al.}(2001)\citenamefont{H.W.C.Postma,
  T.Teepen, Z.Yao, M.Grifoni, and C.Dekker}}]{postma01}
\bibinfo{author}{\bibnamefont{H.W.C.Postma}},
  \bibinfo{author}{\bibnamefont{T.Teepen}},
  \bibinfo{author}{\bibnamefont{Z.Yao}},
  \bibinfo{author}{\bibnamefont{M.Grifoni}}, \bibnamefont{and}
  \bibinfo{author}{\bibnamefont{C.Dekker}}, \bibinfo{journal}{Science}
  \textbf{\bibinfo{volume}{293}}, \bibinfo{pages}{76} (\bibinfo{year}{2001}).

\bibitem[{\citenamefont{J.W.Janssen et~al.}(2002)\citenamefont{J.W.Janssen,
  S.G.Lemay, L.P.Kouwenhoven, and C.Dekker}}]{janssen02}
\bibinfo{author}{\bibnamefont{J.W.Janssen}},
  \bibinfo{author}{\bibnamefont{S.G.Lemay}},
  \bibinfo{author}{\bibnamefont{L.P.Kouwenhoven}}, \bibnamefont{and}
  \bibinfo{author}{\bibnamefont{C.Dekker}}, \bibinfo{journal}{Physical Review
  B} \textbf{\bibinfo{volume}{65}}, \bibinfo{pages}{115423}
  (\bibinfo{year}{2002}).

\bibitem[{\citenamefont{Vitali et~al.}(2004)\citenamefont{Vitali, Burghard,
  M.A.Schneider, Liu, Jayanthi, and Kern}}]{vitali04b}
\bibinfo{author}{\bibfnamefont{L.}~\bibnamefont{Vitali}},
  \bibinfo{author}{\bibfnamefont{M.}~\bibnamefont{Burghard}},
  \bibinfo{author}{\bibnamefont{M.A.Schneider}},
  \bibinfo{author}{\bibfnamefont{L.}~\bibnamefont{Liu}},
  \bibinfo{author}{\bibfnamefont{C.}~\bibnamefont{Jayanthi}}, \bibnamefont{and}
  \bibinfo{author}{\bibfnamefont{K.}~\bibnamefont{Kern}},
  \bibinfo{journal}{Phyiscal Review Letters} \textbf{\bibinfo{volume}{93}},
  \bibinfo{pages}{136103} (\bibinfo{year}{2004}).

\bibitem[{\citenamefont{J.W.G.Wildoer et~al.}(1998)\citenamefont{J.W.G.Wildoer,
  L.C.Venema, A.G.Rinzler, R.E.Smalley, and C.Dekker}}]{wildoer98}
\bibinfo{author}{\bibnamefont{J.W.G.Wildoer}},
  \bibinfo{author}{\bibnamefont{L.C.Venema}},
  \bibinfo{author}{\bibnamefont{A.G.Rinzler}},
  \bibinfo{author}{\bibnamefont{R.E.Smalley}}, \bibnamefont{and}
  \bibinfo{author}{\bibnamefont{C.Dekker}}, \bibinfo{journal}{Nature}
  \textbf{\bibinfo{volume}{391}}, \bibinfo{pages}{59} (\bibinfo{year}{1998}).

\bibitem[{\citenamefont{M.S.Fuhrer et~al.}(2000)\citenamefont{M.S.Fuhrer,
  J.Nygard, L.Shih, M.Forero, Yoon, M.S.C.Mazzoni, Choi, Ihm, S.G.Louie,
  A.Zettl et~al.}}]{fuhrer00}
\bibinfo{author}{\bibnamefont{M.S.Fuhrer}},
  \bibinfo{author}{\bibnamefont{J.Nygard}},
  \bibinfo{author}{\bibnamefont{L.Shih}},
  \bibinfo{author}{\bibnamefont{M.Forero}},
  \bibinfo{author}{\bibfnamefont{Y.-G.} \bibnamefont{Yoon}},
  \bibinfo{author}{\bibnamefont{M.S.C.Mazzoni}},
  \bibinfo{author}{\bibfnamefont{H.~J.} \bibnamefont{Choi}},
  \bibinfo{author}{\bibfnamefont{J.}~\bibnamefont{Ihm}},
  \bibinfo{author}{\bibnamefont{S.G.Louie}},
  \bibinfo{author}{\bibnamefont{A.Zettl}}, \bibnamefont{et~al.},
  \bibinfo{journal}{Science} \textbf{\bibinfo{volume}{288}},
  \bibinfo{pages}{494} (\bibinfo{year}{2000}).

\bibitem[{\citenamefont{J.W.Jassen et~al.}(2002)\citenamefont{J.W.Jassen,
  S.G.Lemay, L.P.Kouwenhoven, and C.Dekker}}]{jassen02}
\bibinfo{author}{\bibnamefont{J.W.Jassen}},
  \bibinfo{author}{\bibnamefont{S.G.Lemay}},
  \bibinfo{author}{\bibnamefont{L.P.Kouwenhoven}}, \bibnamefont{and}
  \bibinfo{author}{\bibnamefont{C.Dekker}}, \bibinfo{journal}{Physical Review
  B} \textbf{\bibinfo{volume}{65}}, \bibinfo{pages}{115423}
  (\bibinfo{year}{2002}).

\bibitem[{\citenamefont{Y.Xue and S.Datta}(1998)}]{xue99}
\bibinfo{author}{\bibnamefont{Y.Xue}} \bibnamefont{and}
  \bibinfo{author}{\bibnamefont{S.Datta}}, \bibinfo{journal}{Physical Review
  Letters} \textbf{\bibinfo{volume}{83}}, \bibinfo{pages}{4844}
  (\bibinfo{year}{1998}).

\bibitem[{\citenamefont{R.Czerw et~al.}(2002)\citenamefont{R.Czerw, B.Foley,
  D.Tekleab, A.Rubio, P.M.Ajayan, and D.L.Carroll}}]{czerw02}
\bibinfo{author}{\bibnamefont{R.Czerw}},
  \bibinfo{author}{\bibnamefont{B.Foley}},
  \bibinfo{author}{\bibnamefont{D.Tekleab}},
  \bibinfo{author}{\bibnamefont{A.Rubio}},
  \bibinfo{author}{\bibnamefont{P.M.Ajayan}}, \bibnamefont{and}
  \bibinfo{author}{\bibnamefont{D.L.Carroll}}, \bibinfo{journal}{Physical
  Review B} \textbf{\bibinfo{volume}{66}}, \bibinfo{pages}{033408}
  (\bibinfo{year}{2002}).

\bibitem[{\citenamefont{F.Léonard and J.Tersoff}(1999)}]{leonard99}
\bibinfo{author}{\bibnamefont{F.Léonard}} \bibnamefont{and}
  \bibinfo{author}{\bibnamefont{J.Tersoff}}, \bibinfo{journal}{Physical Review
  Letters} \textbf{\bibinfo{volume}{83}}, \bibinfo{pages}{5174}
  (\bibinfo{year}{1999}).

\bibitem[{\citenamefont{A.A.Odintsov}(2000)}]{odintsov00}
\bibinfo{author}{\bibnamefont{A.A.Odintsov}}, \bibinfo{journal}{Physical Review
  Letters} \textbf{\bibinfo{volume}{85}}, \bibinfo{pages}{150}
  (\bibinfo{year}{2000}).

\bibitem[{\citenamefont{A.E.Hanna et~al.}(2000)\citenamefont{A.E.Hanna,
  M.T.Tuominen, and M.Tinkham}}]{hanna92}
\bibinfo{author}{\bibnamefont{A.E.Hanna}},
  \bibinfo{author}{\bibnamefont{M.T.Tuominen}}, \bibnamefont{and}
  \bibinfo{author}{\bibnamefont{M.Tinkham}}, \bibinfo{journal}{Physical Review
  Letters} \textbf{\bibinfo{volume}{68}}, \bibinfo{pages}{3228}
  (\bibinfo{year}{2000}).

\bibitem[{\citenamefont{M.S.Dreesselhaus
  et~al.}(2001)\citenamefont{M.S.Dreesselhaus, G.Dresselhaus, and
  P.Avouris}}]{springer80}
\bibinfo{author}{\bibnamefont{M.S.Dreesselhaus}},
  \bibinfo{author}{\bibnamefont{G.Dresselhaus}}, \bibnamefont{and}
  \bibinfo{author}{\bibnamefont{P.Avouris}}, \bibinfo{journal}{Springer Verlag}
  \textbf{\bibinfo{volume}{80}}, \bibinfo{pages}{152} (\bibinfo{year}{2001}).

\bibitem[{\citenamefont{A.Rubio et~al.}(1999)\citenamefont{A.Rubio,
  D.Sánchez-Portal, E.Artacho, P.Ordejón, and J.M.Soler}}]{rubio99}
\bibinfo{author}{\bibnamefont{A.Rubio}},
  \bibinfo{author}{\bibnamefont{D.Sánchez-Portal}},
  \bibinfo{author}{\bibnamefont{E.Artacho}},
  \bibinfo{author}{\bibnamefont{P.Ordejón}}, \bibnamefont{and}
  \bibinfo{author}{\bibnamefont{J.M.Soler}}, \bibinfo{journal}{Physical Review
  Letters} \textbf{\bibinfo{volume}{82}}, \bibinfo{pages}{3520}
  (\bibinfo{year}{1999}).

\bibitem[{\citenamefont{M.Knupfer and H.Peisert}(2004)}]{knupfer04}
\bibinfo{author}{\bibnamefont{M.Knupfer}} \bibnamefont{and}
  \bibinfo{author}{\bibnamefont{H.Peisert}}, \bibinfo{journal}{Physica Status
  solidi (a)} \textbf{\bibinfo{volume}{201}}, \bibinfo{pages}{1055}
  (\bibinfo{year}{2004}).

\bibitem[{\citenamefont{P.Corio et~al.}(2003)\citenamefont{P.Corio, P.S.Santos,
  V.W.Brar, G.G.Samsonidze, S.G.Chou, and M.S.Dresselhaus}}]{corio03}
\bibinfo{author}{\bibnamefont{P.Corio}},
  \bibinfo{author}{\bibnamefont{P.S.Santos}},
  \bibinfo{author}{\bibnamefont{V.W.Brar}},
  \bibinfo{author}{\bibnamefont{G.G.Samsonidze}},
  \bibinfo{author}{\bibnamefont{S.G.Chou}}, \bibnamefont{and}
  \bibinfo{author}{\bibnamefont{M.S.Dresselhaus}}, \bibinfo{journal}{Chemical
  Physics Letters} \textbf{\bibinfo{volume}{370}}, \bibinfo{pages}{675}
  (\bibinfo{year}{2003}).

\bibitem[{\citenamefont{Gupta et~al.}(2004)\citenamefont{Gupta, M.Hughes,
  A.H.Windle, and J.Robertson}}]{gupta04}
\bibinfo{author}{\bibfnamefont{S.}~\bibnamefont{Gupta}},
  \bibinfo{author}{\bibnamefont{M.Hughes}},
  \bibinfo{author}{\bibnamefont{A.H.Windle}}, \bibnamefont{and}
  \bibinfo{author}{\bibnamefont{J.Robertson}}, \bibinfo{journal}{Journal of
  Applied Physics} \textbf{\bibinfo{volume}{95}}, \bibinfo{pages}{2038}
  (\bibinfo{year}{2004}).

\bibitem[{\citenamefont{M.S.Dresselhaus and P.C.Eklund}(2000)}]{dresselhaus00}
\bibinfo{author}{\bibnamefont{M.S.Dresselhaus}} \bibnamefont{and}
  \bibinfo{author}{\bibnamefont{P.C.Eklund}}, \bibinfo{journal}{Advances in
  Physics} \textbf{\bibinfo{volume}{49}}, \bibinfo{pages}{705}
  (\bibinfo{year}{2000}).

\bibitem[{\citenamefont{L.Henrard et~al.}(1999)\citenamefont{L.Henrard,
  E.Hernandez, P.Bernier, and A.Rubio}}]{henrard99}
\bibinfo{author}{\bibnamefont{L.Henrard}},
  \bibinfo{author}{\bibnamefont{E.Hernandez}},
  \bibinfo{author}{\bibnamefont{P.Bernier}}, \bibnamefont{and}
  \bibinfo{author}{\bibnamefont{A.Rubio}}, \bibinfo{journal}{Physical Review B}
  \textbf{\bibinfo{volume}{60}}, \bibinfo{pages}{8521} (\bibinfo{year}{1999}).

\bibitem[{\citenamefont{S.M.Sharma et~al.}(2001)\citenamefont{S.M.Sharma,
  S.Karmakar, S.K.Sikka, P.V.Teredesai, A.K.Sood, A.Govindaraj, and
  C.N.R.Rao}}]{sharma01}
\bibinfo{author}{\bibnamefont{S.M.Sharma}},
  \bibinfo{author}{\bibnamefont{S.Karmakar}},
  \bibinfo{author}{\bibnamefont{S.K.Sikka}},
  \bibinfo{author}{\bibnamefont{P.V.Teredesai}},
  \bibinfo{author}{\bibnamefont{A.K.Sood}},
  \bibinfo{author}{\bibnamefont{A.Govindaraj}}, \bibnamefont{and}
  \bibinfo{author}{\bibnamefont{C.N.R.Rao}}, \bibinfo{journal}{Physical Review
  B} \textbf{\bibinfo{volume}{63}}, \bibinfo{pages}{205417}
  (\bibinfo{year}{2001}).

\bibitem[{\citenamefont{I.Loa}(2003)}]{loa03}
\bibinfo{author}{\bibnamefont{I.Loa}}, \bibinfo{journal}{Journal of Raman
  Spectroscopy} \textbf{\bibinfo{volume}{34}}, \bibinfo{pages}{611}
  (\bibinfo{year}{2003}).

\end{thebibliography}

\end{document}